\documentclass[english,english,aps,prd,amsmath,amssymb,superscriptaddress]{revtex4}
\usepackage[T1]{fontenc}
\usepackage[latin9]{inputenc}
\usepackage{amsmath}
\usepackage{graphicx}
\usepackage{amssymb}
\usepackage{esint}

\makeatletter
\@ifundefined{textcolor}{}
{%
 \definecolor{BLACK}{gray}{0}
 \definecolor{WHITE}{gray}{1}
 \definecolor{RED}{rgb}{1,0,0}
 \definecolor{GREEN}{rgb}{0,1,0}
 \definecolor{BLUE}{rgb}{0,0,1}
 \definecolor{CYAN}{cmyk}{1,0,0,0}
 \definecolor{MAGENTA}{cmyk}{0,1,0,0}
 \definecolor{YELLOW}{cmyk}{0,0,1,0}
 }

\usepackage{slashed}
\usepackage{babel}
\usepackage{bbm}

\makeatother

\usepackage{babel}

\begin{document}

\title{Refractive Index of Light in the Quark-Gluon Plasma with the Hard-Thermal-Loop Perturbation Theory}

\author{Juan Liu}

\author{M.J. Luo}

\author{Qun Wang}

\author{Hao-jie Xu}

\affiliation{Interdisciplinary Center for Theoretical Study and Department of
Modern Physics, University of Science and Technology of China, Anhui
230026, People's Republic of China}
\begin{abstract}
The electric permittivity and magnetic permeability for the quark-gluon plasma (QGP) is calculated within the hard-thermal-loop (HTL)
perturbation theory. The refractive indices in the magnetizable and
nonmagnetizable plasmas are calculated. In a magnetizable plasma,
there is a frequency pole $\omega_{mp}$ in the magnetic permeability
and the refractive index. The refractive index becomes negative in
the range $\omega\in[k,\omega_{mp}]$, where $k$ is the wave number,
but no propagating modes are found. In a non-magnetizable plasma,
the magnetic permeability and the refractive index are always positive.
This marks the main distinction of a weakly coupled plasma from a
strongly coupled one, where the negative refraction is shown to exist in a
holographic theory. 
\end{abstract}
\maketitle

\section{Introduction}

The quark-gluon plasma (QGP) is believed to be a new state of matter
of the strong interaction produced in ultrarelativistic heavy ion
collisions (see, e.g., \cite{Rischke:2003mt,Gyulassy:2004zy,Jacobs:2004qv,Adams:2005dq}
for reviews). The QGP is a plasma containing electrically charged quarks;
its electromagnetic property is an important aspect of its nature.
Among all observables for the QGP, the high-energy photons may provide
clean probes to hot and dense medium \cite{McLerran:1984ay,Kapusta:1991qp,Alam:2000bu,Arnold:2001ba,Arnold:2001ms,Adler:2005ig,Abelev:2009hx,Prasad:2011mx}.
Although the electromagnetic nature gives us an impression that the
interaction between low-energy photons and the medium would be weak,
it could be more significant than we previously thought due to extremely
high temperature and density of the QGP. So the optical properties
of the QGP are not a trivial issue. 

One of the most important optical properties is the refractive index
(RI), which measures the speed of light in a medium relative to vacuum.
The negative refraction (NR) is a very interesting phenomenon of materials
which was first theoretically proposed by Veselago in 1968 \cite{Veselago:1968}.
The physical nature of such a property is that the electromagnetic
phase velocity is in the opposite direction to the energy flow. The NR
leads to many interesting phenomena in materials such as the modified
fraction law \cite{Shelby:2001,Smith:2004} and bremsstrahlung radiation,
the reverse Doppler shift \cite{PhysRevLett.91.133901} and Cherenkov
radiation, etc.. The most striking application of the NR materials
is the optical cloak \cite{Schurig:2006}, an attractive topic in
science fiction. Recently, the strongly coupled plasma has been found to
have negative refraction, which was first proposed in Ref. \cite{Amariti:2010jw} and afterwards followed in Refs. \cite{Ge:2010yc,Gao:2010ie,Bigazzi:2011it,Amariti:2011dm} in holographic model. Furthermore, it was proven to be a possible generic phenomenon in charged hydrodynamical systems \cite{Amariti:2011dj}. 

Motivated by the above finding in the strongly coupled plasma, in
this paper we will calculate within the hard-thermal-loop (HTL) perturbation
theory the RI via the electric permittivity $\epsilon$ and the magnetic
permeability $\mu$. Since a QGP is composed of electrically charged
quarks instead of magnetic monopoles, the electric and magnetic sector
do not play equal roles. We will show that the physical definition
and behavior of $\mu$ (and then the RI) can be very different due
to specific magnetic response of the QGP. Therefore a plasma can be
classified into two types: magnetizable and nonmagnetizable. In a
magnetizable plasma the magnetization is realistic, while in a nonmagnetizable
plasma it does not make physical sense any more. We will calculate
the RI and analyze their properties in these two types of plasmas.
If the QGP is magnetizable, we will show that there is a frequency
pole $\omega_{mp}$ in $\mu$ and then the RI, leading to the NRI
in the range $\omega\in[k,\omega_{mp}]$, where $k$ is the wave number,
but there are no propagating modes in the NRI region. In a nonmagnetizable
plasma, $\mu$ and the RI are always positive. 

Our results are different from the holographic treatment in following
respects. First, in our perturbative treatment with the HTL, the QGP
is magnetizable in the frequency range where the NR takes place. The
definition of the magnetic permeability is normal and the magnetization
density makes physical sense in this region. In contrast, the plasma
considered in the holographic theory \cite{Amariti:2010jw,Ge:2010yc,Gao:2010ie,Bigazzi:2011it,Amariti:2011dm}
is nonmagnetizable or strongly dielectric, therefore the Landau-Lifshits
description \cite{Landau:1984} of the magnetic permeability has to
be applied. Second, the dispersion relation with the NRI behaves differently.
There is a frequency pole below $\omega_{p}$ for $\mu$ and the RI
in our approach, but there is no such a singularity in the holographic
treatment \cite{Amariti:2010jw,Ge:2010yc,Gao:2010ie,Bigazzi:2011it,Amariti:2011dm}. Finally
we fail to find any propagating modes with the NRI within the HTL
perturbation theory in both magnetizable and nonmagnetizable cases.
This marks the main distinction from a strongly coupled plasma where
the NR is shown to exist in the holographic theory. 

The paper is organized as follows. In Sec. \ref{sec:em-wave} we
briefly review the formalism of computing electromagnetic properties
in an isotropic and anisotropy medium. In Sec. \ref{sec:htl}, we
calculate the RI from the HTL self-energy of photons in magnetizable
and nonmagnetizable plasmas. In Sec. \ref{sec:anisotropy}, we generalize
our calculation to an anisotropic QGP. We finally present the summary
and conclusion in Sec. \ref{sec:conclusion}. We take $k_{B}=\hbar=c=1$
and $g_{\mu\nu}=(+,-,-,-)$ for convention. Here is a summary of abbreviations:
quark-gluon plasma (QGP), hard-thermal-loop (HTL), refractive index
(RI), negative refraction (NR), negative refraction index (NRI).

\section{Propagation of electromagnetic wave in medium}

\label{sec:em-wave}The description of electromagnetic waves propagating
in a continuous medium is normally given by the electric and magnetic
field $E$ and $B$, and the macroscopic field $D$ and $H$. Their relations
give the definition of the electric permittivity and the magnetic
permeability \begin{eqnarray}
D_{i} & = & \epsilon_{ij}(\omega,k)E_{j},\nonumber \\
B_{i} & = & \mu_{ij}(\omega,k)H_{j},\label{eq:definition_epsilon_mu}\end{eqnarray}
where $i,j=1,2,3$ are spatial indices, and $\omega$ and $k\equiv|\mathbf{k}|$
are the frequency and wave number, respectively. This scenario applies
to magnetizable materials. 

Another description was proposed by Landau and Lifshits: that the magnetization
$\mathbf{M}$ loses its usual physical meaning as a magnetic moment
density, and so does the magnetic permeability $\mu(\omega,k)$, when
the characteristic electromagnetic wavelength $\lambda$ is large
enough to violate $\lambda^{2}\ll\chi c^{2}/\omega^{2}$, where $\chi$
is the magnetic susceptibility. This scenario applies to nonmagnetizable
materials. In this case, $E$, $B$ and $D$ are proper quantities
with $\mu(\omega,k)$ being set to unity \cite{Landau:1984}. The
electromagnetic properties of a medium can be provided by \begin{equation}
D_{i}=\tilde{\epsilon}_{ij}(\omega,k)E_{j},\label{eq:tensor_epsilon}\end{equation}
where $\tilde{\epsilon}(\omega,k)$ is a generalized electric permittivity
encoding all electromagnetic response of the medium, in replacement
of $\epsilon(\omega,k)$ and $\mu(\omega,k)$ in the conventional
case. One can obtain the effective electric permittivity $\epsilon(\omega)$
and magnetic permeability $\mu(\omega)$ by expanding $\tilde{\epsilon}(\omega,k)$
in powers of $k^{2}$ \cite{PhysRevB.69.165112} \begin{eqnarray}
\tilde{\epsilon}(\omega,k) & = & \epsilon(\omega)+\frac{k^{2}}{\omega^{2}}\left[1-\frac{1}{\mu(\omega)}\right]+\mathcal{O}(k^{4}).\label{eq:epsilon_expansion}\end{eqnarray}
Note that $\epsilon(\omega)$ and $\mu(\omega)$ only depend on $\omega$,
and $\mu(\omega)$ comes from the dielectric part of $\tilde{\epsilon}(\omega,k)$
which is different from the conventional definition of the magnetic
permeability, especially at low frequency. Only at high frequency,
since the magnetic response cannot follow the fast variation of the
electromagnetic wave, do these two scenarios coincide and give the vacuum
value. 

To describe the electromagnetic properties of the QGP covariantly,
it is natural to use the fluid four-velocity $u^{\alpha}$ to define
the electric and magnetic field strength \begin{equation}
\tilde{E}^{\mu}=u_{\alpha}F^{\mu\alpha},\;\tilde{B}^{\mu}=\frac{1}{2}\epsilon^{\mu\rho\alpha\beta}u_{\rho}F_{\alpha\beta},\end{equation}
where $\epsilon_{\mu\nu\alpha\beta}=-\epsilon^{\mu\nu\alpha\beta}=-1,1$
for that the order of Lorentz indices $(\mu\nu\alpha\beta)$ is an
even/odd permutation of $(0123)$. These quantities are often used
when we consider the interaction with a plasma. Then we can immediately
write $F^{\mu\nu}$ as \begin{equation}
F^{\mu\nu}=\tilde{E}^{\mu}u^{\nu}-\tilde{E}^{\nu}u^{\mu}-\epsilon^{\mu\nu\alpha\beta}\tilde{B}_{\alpha}u_{\beta}.\end{equation}
The free action can be expressed in terms of Fourier transformed $\tilde{E}$
and $\tilde{B}$ as \begin{equation}
S_{0}=-\frac{1}{2}\int\frac{d^{4}K}{(2\pi)^{4}}\left[\tilde{E}^{\mu}(K)\tilde{E}_{\mu}(-K)-\tilde{B}^{\mu}(K)\tilde{B}_{\mu}(-K)\right].\end{equation}
Including the medium effect, the action becomes \begin{eqnarray}
S & = & S_{0}-\frac{1}{2}\int\frac{d^{4}K}{(2\pi)^{4}}A^{\mu}(K)\Pi_{\mu\nu}(K)A^{\nu}(-K)+..,\end{eqnarray}
where $A^{\mu}(K)$ is the photon field in momentum space. The medium
effect is characterized by the photon self-energy $\Pi_{\mu\nu}(K)$.
One can then extract the electric permittivity and the magnetic permeability
from the action $S$.

\subsection{Isotropic medium}

The self-energy must satisfy the Ward identity \begin{equation}
K^{\mu}\Pi_{\mu\nu}(K)=0,\label{eq:ward}\end{equation}
in which $K_{\mu}=(\omega,k_{i})$ is a 4-momentum. In an isotropic
medium, the general solution of Eq.(\ref{eq:ward}) for $\Pi_{\mu\nu}(K)$
is a linear combination of available symmetric tensors \begin{equation}
\{g_{\mu\nu},\; K_{\mu}K_{\nu},\; u_{\mu}u_{\nu},\; u_{\mu}K_{\nu}+u_{\nu}K_{\mu}\}.\label{eq:tensor-basis-1}\end{equation}
Two independent solutions are the transverse and longitudinal projectors
$P_{\mu\nu}$ and $Q_{\mu\nu}$,

\begin{eqnarray}
P_{\mu\nu} & = & g_{\mu\nu}-u_{\mu}u_{\nu}+\frac{1}{k^{2}}\left(K_{\mu}-\omega u_{\mu}\right)\left(K_{\nu}-\omega u_{\nu}\right),\nonumber \\
Q_{\mu\nu} & = & \frac{-1}{K^{2}k^{2}}\left(\omega K_{\mu}-K^{2}u_{\mu}\right)\left(\omega K_{\nu}-K^{2}u_{\nu}\right).\label{eq:projectors}\end{eqnarray}
Therefore $\Pi_{\mu\nu}(K)$ can be expanded in these projectors as
\begin{equation}
\Pi_{\mu\nu}(K)=\Pi_{T}(K)P_{\mu\nu}+\Pi_{L}(K)Q_{\mu\nu},\label{eq:PQ}\end{equation}
where $\Pi_{T/L}(K)$ are transverse/longitudinal part of the self-energy.
The full inverse propagator for the photon is \begin{eqnarray}
D_{\mu\nu}^{-1} & = & D_{(0)\mu\nu}^{-1}+\Pi_{\mu\nu}=(-K^{2}+\Pi_{L})Q_{\mu\nu}+(-K^{2}+\Pi_{T})P_{\mu\nu}-\frac{1}{1-\eta}K_{\mu}K_{\nu}\end{eqnarray}
where the last term is the gauge-fixing term and $\eta$ is the gauge
parameter. Here we choose the covariant gauge $K\cdot A_{K}=0$ so
the gauge-fixing term does not appear in the action. 

In magnetizable case, the action can then be evaluated with $D_{\mu\nu}^{-1}(K)$
as \begin{eqnarray}
S & = & \frac{1}{2}\int d^{4}KA_{\mu}(K)\left(D^{-1}\right)^{\mu\nu}A_{\nu}(-K)\nonumber \\
 & = & -\frac{1}{2}\int d^{4}K\left[\epsilon_{\mu\nu}\tilde{E}^{\mu}(K)\tilde{E}^{\nu}(-K)-\frac{1}{\mu}\tilde{B}_{\mu}(K)\tilde{B}^{\mu}(-K)\right],\label{eq:full_action_weldon}\end{eqnarray}
where\begin{eqnarray}
\epsilon_{\mu\nu}(\omega,k) & = & \epsilon(\omega,k)g_{\mu\nu}=\left(1-\frac{\Pi_{L}}{K^{2}}\right)g_{\mu\nu},\nonumber \\
\frac{1}{\mu(\omega,k)} & = & 1+\frac{K^{2}\Pi_{T}-\omega^{2}\Pi_{L}}{k^{2}K^{2}}.\label{eq:magnetizable_epsilon_mu}\end{eqnarray}

In the nonmagnetizable case or the Landau-Lifshits scenario, we obtain
\begin{equation}
S=-\frac{1}{2}\int d^{4}K\left[\tilde{\epsilon}\tilde{E}_{\mu}(K)\tilde{E}^{\mu}(-K)-\tilde{B}_{\mu}(K)\tilde{B}^{\mu}(-K)\right],\label{eq:un-mag}\end{equation}
where \begin{equation}
\tilde{\epsilon}(\omega,k)=1-\frac{\Pi_{T}}{\omega^{2}}.\label{eq:epsilon-t1}\end{equation}
Then the effective electric permittivity and magnetic permeability
can be extracted from $\tilde{\epsilon}(\omega,k)$ \begin{eqnarray}
\epsilon(\omega) & = & 1-\frac{\Pi_{T}^{(0)}(\omega)}{\omega^{2}},\nonumber \\
\frac{1}{\mu(\omega)} & = & 1+\Pi_{T}^{(2)}(\omega),\label{eq:unmagnetizable_epsilon_mu}\end{eqnarray}
where we have expanded the $\Pi_{T}$ in powers of $k$ by  \begin{equation}
\Pi_{T}(\omega,k)=\Pi_{T}^{(0)}(\omega)+k^{2}\Pi_{T}^{(2)}(\omega)+\mathcal{O}(k^{4}).\label{eq:epsilon-t3}\end{equation}

\subsection{Anisotropic medium}

Suppose there is one special direction $r_{\mu}=(0,\mathbf{r})$ in
a most simple anisotropic medium. Now there are three vectors $\{\tilde{r}_{\mu},K_{\mu},u_{\mu}\}$
out of which the tensorial bases for $\Pi_{\mu\nu}$ are composed,
where $\tilde{r}_{\mu}\equiv P_{\mu\nu}r^{\nu}$ is a vector normal
to $K^{\mu}$ with $P_{\mu\nu}$ defined by Eq. (\ref{eq:projectors}).
The available symmetric tensors have additional elements besides those
in (\ref{eq:tensor-basis-1}), \begin{equation}
\{...,\;\tilde{r}_{\mu}\tilde{r}_{\nu},\;\tilde{r}_{\mu}K_{\nu}+\tilde{r}_{\nu}K_{\mu},\;\tilde{r}_{\mu}u_{\nu}+\tilde{r}_{\nu}u_{\mu}\}.\label{eq:basis-2}\end{equation}
As a consequence, the tensorial bases that satisfy the Ward identity
(\ref{eq:ward}) now become \begin{equation}
\{P_{\mu\nu},Q_{\mu\nu},C_{\mu\nu},G_{\mu\nu}\},\end{equation}
where two extra projectors are given by\begin{eqnarray}
C_{\mu\nu} & = & \frac{\tilde{r}_{\mu}\tilde{r}_{\nu}}{\tilde{r}^{2}},\nonumber \\
G_{\mu\nu} & = & \left(K_{\mu}\tilde{r}_{\nu}+K_{\nu}\tilde{r}_{\mu}\right)-\frac{K^{2}}{\omega}\left(\tilde{r}_{\mu}u_{\nu}+\tilde{r}_{\nu}u_{\mu}\right),\end{eqnarray}
which obey $K^{\mu}C_{\mu\nu}=K^{\mu}G_{\mu\nu}=0$. Then the self-energy
tensor can be expanded as \begin{equation}
\Pi_{\mu\nu}=\alpha P_{\mu\nu}+\beta Q_{\mu\nu}+\gamma C_{\mu\nu}+\delta G_{\mu\nu},\label{eq:PQCD}\end{equation}
where $\alpha,\beta,\gamma,\delta$ are structure functions. Inserting
the above into the full propagator inverse $D_{\mu\nu}^{-1}(K)$ one
obtains the full action (\ref{eq:full_action_weldon}) with \begin{eqnarray}
\epsilon_{\mu\nu} & = & \left(1-\frac{\beta}{K^{2}}\right)g_{\mu\nu}-\frac{1}{\omega^{2}}\left[\gamma\frac{\tilde{r}_{\mu}\tilde{r}_{\nu}}{\tilde{r}^{2}}+\delta\left(K_{\mu}\tilde{r}_{\nu}+K_{\nu}\tilde{r}_{\mu}\right)\right],\nonumber \\
\frac{1}{\mu} & = & 1+\frac{K^{2}\alpha-\omega^{2}\beta}{k^{2}K^{2}}.\end{eqnarray}
The transverse component of the action (\ref{eq:full_action_weldon})
gives \begin{eqnarray}
S_{T} & = & -\frac{1}{2}\int d^{4}K\left[\epsilon_{\mu\nu}^{T}\tilde{E}_{T}^{\mu}(K)\tilde{E_{T}^{\nu}}(-K)-\frac{1}{\mu}\tilde{B}_{\mu}(K)\tilde{B}^{\mu}(-K)\right],\end{eqnarray}
where $\tilde{E}_{T}^{\mu}(K)=P^{\mu\nu}\tilde{E}_{T,\nu}(K)$. Note
that the magnetic part is always transverse. $\epsilon_{\mu\nu}^{T}$
is the transverse projection of $\epsilon_{\mu\nu}$ and is given by

\begin{eqnarray}
\epsilon_{\mu\nu}^{T} & = & \left(1-\frac{\beta}{K^{2}}\right)g_{\mu\nu}-\frac{\gamma}{\omega^{2}}C_{\mu\nu},\end{eqnarray}
Similarly, in the nonmagnetizable case, the transverse part of the action
is

\begin{equation}
S_{T}=-\frac{1}{2}\int d^{4}K\left[\tilde{\epsilon}_{\mu\nu}^{T}\tilde{E}_{T}^{\mu}(K)\tilde{E}_{T}^{\nu}(-K)-\tilde{B}_{\mu}(K)\tilde{B}^{\mu}(-K)\right],\end{equation}
where

\begin{equation}
\tilde{\epsilon}_{\mu\nu}^{T}=\left(1-\frac{\alpha}{\omega^{2}}\right)g_{\mu\nu}-\frac{\gamma}{\omega^{2}}C_{\mu\nu}.\end{equation}

One has to diagonalize $\epsilon_{\mu\nu}^{T}$ (magnetizable case)
and $\tilde{\epsilon}_{\mu\nu}^{T}$ (nonmagnetizable case) in order
to obtain the eigenvalues of the electric permittivity for the left-
and right-handed polarized photons. Let us consider an analytically
solvable case as follows. In a rest frame with the fluid velocity
$u^{\mu}=(1,0,0,0)$, since $\tilde{E}^{0}=0$, so the spatial part
of $\epsilon_{\mu\nu}^{T}$ or $\tilde{\epsilon}_{\mu\nu}^{T}$, a
$3\times3$ matrix $\epsilon_{ij}^{T}$ or $\tilde{\epsilon}_{ij}^{T}$
for $i,j=1,2,3$, is

\[
\epsilon_{ij}^{T}=\left(1-\frac{\beta}{K^{2}}\right)\delta_{ij}-\frac{\gamma}{\omega^{2}}\frac{\mathbf{r}_{Ti}\mathbf{r}_{Tj}}{\mathbf{r}_{T}^{2}},\]
and\[
\tilde{\epsilon}_{ij}^{T}=\left(1-\frac{\alpha}{\omega^{2}}\right)\delta_{ij}-\frac{\gamma}{\omega^{2}}\frac{\mathbf{r}_{Ti}\mathbf{r}_{Tj}}{\mathbf{r}_{T}^{2}},\]
where we have defined $\mathbf{r}_{T}=\mathbf{r}-(\mathbf{r}\cdot\hat{\mathbf{k}})\hat{\mathbf{k}}$
and used $\tilde{r}^{2}=-\tilde{\mathbf{r}}_{T}^{2}$. One obtains
the eigenvalues for the left- and right-handed polarized photons,
for a magnetizable plasma,\begin{equation}
\epsilon_{L}=1-\frac{\beta}{K^{2}},\;\epsilon_{R}=1-\frac{\beta}{K^{2}}-\frac{\gamma}{\omega^{2}},\label{eq:epsilon_mag}\end{equation}
and those for an nonmagnetizable plasma, 

\begin{equation}
\tilde{\epsilon}_{L}=1-\frac{\alpha}{\omega^{2}},\;\tilde{\epsilon}_{R}=1-\frac{\alpha}{\omega^{2}}-\frac{\gamma}{\omega^{2}}.\label{eq:epsilon_unmag}\end{equation}

\subsection{Refractive Index}

The refractive index is normally defined by $n^{2}=\epsilon\mu$,
but the quadratic nature of such a definition implies that it is not
sensitive to the sign of $\epsilon$ and $\mu$. It is known that
the sign change of $\epsilon$ and $\mu$ corresponds to a cross over
between different branches of the square root, from $n=\sqrt{\epsilon\mu}$
to $n=-\sqrt{\epsilon\mu}$, or from the positive refractive index
to the negative one. We can see in the following that the sign of
$\epsilon$ and $\mu$ have a significant physical implication. The
phase velocity is defined by \begin{equation}
\mathbf{v}_{p}=\frac{1}{\mathrm{Re}(n)}\hat{\mathbf{k}}=v_{p}\hat{\mathbf{k}},\end{equation}
whose sign is the same as that of $\mathrm{Re}(n)$. But the direction
of the energy flow or the Poynting vector is not affected by the sign
of $\epsilon$ and $\mu$. In a medium with small dissipation, the
direction of the energy flow coincides with that of the group velocity,
\begin{equation}
\mathbf{S}=v_{g}U\hat{\mathbf{k}},\end{equation}
where $U$ is a positive time-averaged energy density and $v_{g}=d\omega/dk$. 

So the direction of the phase velocity can be opposite to the energy
flow or the group velocity if we have, e.g., a negative phase velocity
and a positive group one or vice versa \begin{equation}
v_{p}<0,\; v_{g}>0.\end{equation}
This criterion of the antiparallelism for the phase velocity and
the energy flow is equivalent to a better definition called the Depine-Lakhtakia
(DL) index \cite{DL}, \begin{equation}
n_{DL}=\left|\epsilon\right|\mathrm{Re}(\mu)+\left|\mu\right|\mathrm{Re}(\epsilon).\label{eq:nDL}\end{equation}
When $n_{DL}<0$, the directions of the phase velocity and the energy
flow are opposite. So $n_{DL}$ is a good quantity for covering the
NRI. We will calculate both $n_{DL}$ and $n$ in the next section.

\section{Hard-Thermal-Loop self-energy for photon }

\label{sec:htl}
The self-energy tensor of photon in a plasma can be calculated by the standard perturbative technique of Feynman diagram at finite temperature and density (i.e., a finite chemical potential). However, a complete calculation of $\Pi_{\mu\nu}$ is rather involved because of the significantly high temperature of the QGP, even at one loop level in which the self-energy is given by the exchange of quark loops, so we only limit ourselves to the high-temperature approximation, which means that the temperature is much larger than the quark mass and external momenta, named the HTL part of $\Pi_{\mu\nu}$. In this section, we will investigate the refractive index for the HTL self-energy $\Pi_{\mu\nu}$
of the photon from the quark loops in the QGP \cite{Weldon:1982aq,Pisarski:1988vd,Bellac:1996}.
The HTL approximation works well at high temperature where the QGP is thought to be weakly coupled. The HTL self-energy reads

\begin{eqnarray}
\Pi_{T}(\omega,k) & = & \frac{1}{2}m_{D}^{2}\left[\frac{\omega^{2}}{k^{2}}+\left(1-\frac{\omega^{2}}{k^{2}}\right)\frac{\omega}{2k}\log\frac{\omega+k}{\omega-k}\right],\nonumber \\
\Pi_{L}(\omega,k) & = & m_{D}^{2}\left(1-\frac{\omega^{2}}{k^{2}}\right)\left[1-\frac{\omega}{2k}\log\frac{\omega+k}{\omega-k}\right],\label{eq:HTL}\end{eqnarray}
in which the Debye mass squared is \begin{equation}
m_{D}^{2}\equiv e^{2}N_{c}\sum^{N_f}_{f}\left(\frac{1}{3}T^{2}+\frac{\mu_{f}^{2}}{\pi^{2}}\right)Q_{f}^{2},\end{equation}
where $e$ is the electric charge, $\mu_{f}$ and $Q_{f}$ are the
quark chemical potential and the electric charge for the flavor species
$f$, $N_{c}$ and $N_{f}$ are the number of colors
and flavors, respectively. When $\omega>k$, $\Pi_{T,L}(\omega,k)$
are real, meaning that the medium has no dissipation for propagating
modes. When $k>\omega$, the imaginary parts appear, the medium becomes
dissipative due to the Landau damping effect.

\subsection{Magnetizable Plasma}

\begin{figure}
\includegraphics[scale=0.5]{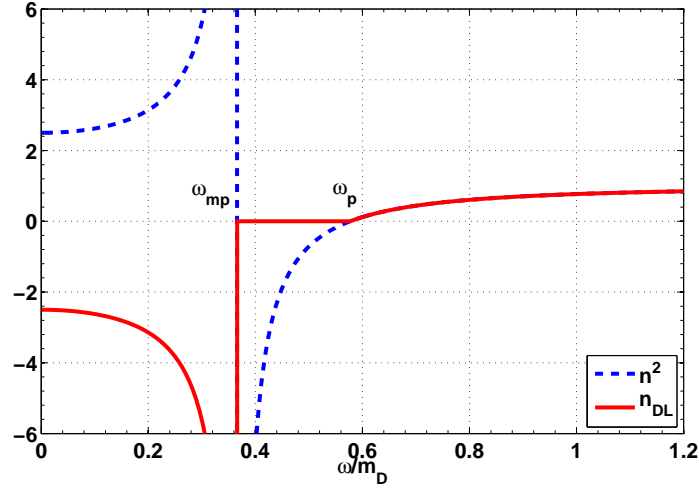}

\caption{\label{fig:ri-m-s}The RI for a magnetizable plasma in small $k$
expansion: the conventional refractive index $n^{2}$ and the Depine-Lakhtakia
index $n_{DL}$ as functions of $\omega$. We take $m_{D}$ as the
unit for $k$ and $\omega$. }

\end{figure}

A magnetizable plasma is the one in which the magnetic moment and the
magnetic permeability $\mu(\omega,k)$ have ordinary physical meanings.
Substituting the self-energy (\ref{eq:HTL}) into Eq. (\ref{eq:magnetizable_epsilon_mu})
we obtain 

\begin{eqnarray}
\epsilon(\omega,k) & = & 1+\frac{m_{D}^{2}}{k^{2}}\left(1-\frac{\omega}{2k}\log\frac{\omega+k}{\omega-k}\right),\nonumber \\
\mu(\omega,k) & = & \frac{4k^{4}}{4k^{4}+6m_{D}^{2}\omega^{2}+m_{D}^{2}(k^{2}-3\omega^{2})\frac{\omega}{k}\log\frac{\omega+k}{\omega-k}}.\label{eq:whole_dispersion_mag}\end{eqnarray}
In principle, the plasma contains not only temporal dispersion, but
also spatial dispersion. In an isotropic medium, if the phase velocity
of light is much larger than the thermal velocity of plasma particles,
the spatial dispersion is small, and then $\epsilon(\omega,k)$ and
$\mu(\omega,k)$ can be assumed to be independent of $k$. In this
small $k$ limit, we have $\omega>k$, both $\epsilon(\omega,k)$
and $\mu(\omega,k)$ are real. This is equivalent to expanding $\epsilon(\omega,k)$
and $\mu(\omega,k)$ in $k$ around $k=0$ \begin{eqnarray}
\epsilon(\omega) & \approx & 1-\frac{m_{D}^{2}}{3\omega^{2}}+\mathcal{O}(k^{2}),\nonumber \\
\mu(\omega) & \approx & \frac{1}{1-\frac{2m_{D}^{2}}{15\omega^{2}}}+\mathcal{O}(k^{2}).\label{eq:epsilon_mu_HTL_mag}\end{eqnarray}
The first observation of Eq. (\ref{eq:epsilon_mu_HTL_mag}) is that
$\epsilon(\omega)$ is negative for $\omega<\omega_{p}=\sqrt{1/3}m_{D}$,
where $\omega_{p}$ is called the Debye screening frequency. The second
observation is: there is a pole at $\omega_{mp}=\sqrt{2/15}m_{D}$
in $\mu(\omega)$, below which both $\epsilon(\omega)$ and $\mu(\omega)$
become negative. We show $n^{2}$ and $n_{DL}$ as functions of $\omega/m_{D}$
in Fig. \ref{fig:ri-m-s}. At high frequencies, the refractive index
is always less than unity, meaning that the phase velocity is greater
than the speed of light. There is a frequency gap between $\omega_{mp}$
and $\omega_{p}$ where $n^{2}<0$ and $n_{DL}=0$. For frequencies
in the gap, the light cannot propagate or the plasma is opaque to
electromagnetic waves. The width of the gap is proportional to $m_{D}$,
i.e., the higher the temperatures and/or densities, the broader the
gap is. At frequencies lower than the pole, $k<\omega<\omega_{mp}$,
$n_{DL}$ becomes negative. However, the propagating modes should satisfy
the transverse dispersion relation $n^{2}\omega^{2}=k^{2}$; it has
no solution in the range $\omega\in[k,\omega_{mp}]$, indicating that
there are no propagating modes in the NRI region. Note that $\mu(\omega)$
and $n^{2}$ diverge at $\omega_{mp}$, where the magnetization is
large and resonantly oscillates with the electromagnetic wave. The
phase and group velocity are small in the region and approaches zero
at $\omega_{mp}$. In the range $\omega\in[\omega_{mp},\omega_{p}]$,
we see $n^{2}<0$, i.e., the RI is purely imaginary and the electromagnetic
wave is damped. 

\begin{figure}
\includegraphics[scale=0.5]{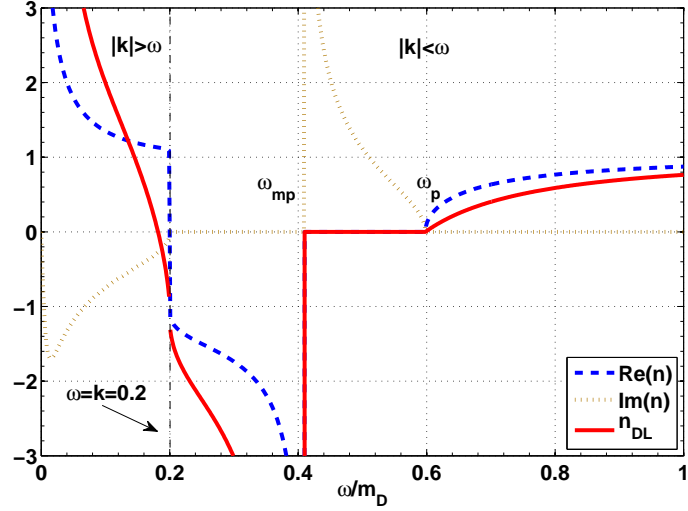}

\caption{\label{fig:ri-m}The refractive indices $n$ (real part: blue dashed,
imaginary part: brown dotted) and $n_{DL}$ (red solid) for a magnetizable
plasma at $k=0.2$. In the region $\omega<k=0.2$, $n$ is complex
and has real and imaginary parts, meaning that the medium is dissipative.
In the range $k<\omega<\omega_{mp}$, $n$ is purely real and negative,
$n<0$, but there is no propagating mode which satisfies the dispersion
$n^{2}\omega^{2}=k^{2}$. In the range $\omega_{mp}<\omega<\omega_{p}$,
$n$ becomes purely imaginary, so the medium is opaque. When $\omega_{p}<\omega$,
the real part of $n$ is positive, $\mathrm{Re}(n)>0$. We take $m_{D}$
as the unit for $k$ and $\omega$. }

\end{figure}

\begin{figure}
\includegraphics[scale=0.5]{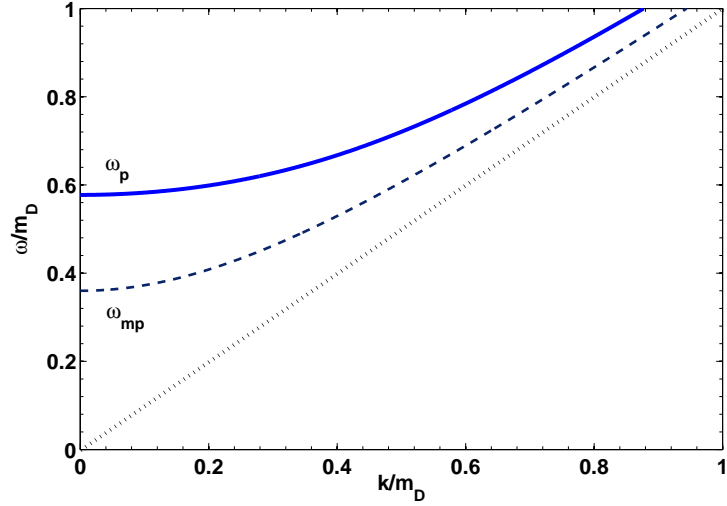}

\caption{\label{fig:disp-m}The dispersion relation $\omega(k)$ for magnetizable
and nonmagnetizable plasmas (blue solid line). The dotted line is the
light cone $\omega=k$. If the plasma is magnetizable, we have poles
on the dashed line, but there are no poles in nonmagnetizable plasma.
We take $m_{D}$ as the unit for $k$ and $\omega$.}

\end{figure}

Now we can go beyond the small $k$ limit by working with the full
version of $\epsilon$ and $\mu$ in Eq. (\ref{eq:whole_dispersion_mag}).
We show $n$ and $n_{DL}$ at a fixed value $k/m_{D}=0.2$ in Fig.
\ref{fig:ri-m}. The values of $\omega_{mp}$ and $\omega_{p}$ increases
with $k$. There is a jump in $n_{DL}$ and $\mathrm{Re}(n)$ at $\omega=k$,
they change from positive to negative values from $\omega<k$ to $\omega>k$.
When $\omega<k$, both $\epsilon$ and $\mu$ become complex, indicating
that the medium is dissipative. When $k<\omega<\omega_{mp}$, we have
$n_{DL}<0$ and $\mathrm{Re}(n)<0$, but no propagating modes are
found since the dispersion $n^{2}\omega^{2}=k^{2}$ cannot hold in
this region. In the frequency range, $\omega_{mp}<\omega<\omega_{p}$,
the refractive index is purely imaginary, so any propagating modes
are forbidden. When $\omega_{p}<\omega$, there are normal propagating
modes with positive refraction. The dispersion relation $\omega(k)$
is shown in Fig. \ref{fig:disp-m}. 

\subsection{Nonmagnetizable plasma}

\begin{figure}
\includegraphics[scale=0.5]{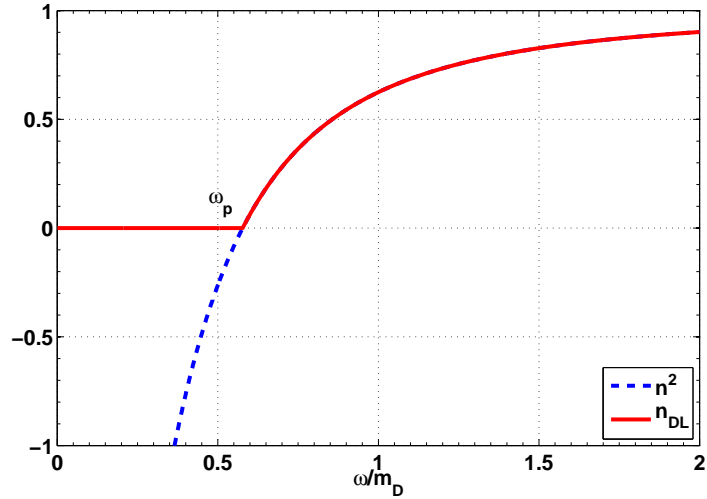}

\caption{\label{fig:ri-unm}The RI for the nonmagnetizable plasma by the conventional
refractive index $n^{2}$ and the Depine-Lakhtakia index $n_{DL}$ as
functions of $\omega$. We take $m_{D}$ as the unit for $k$ and
$\omega$.}

\end{figure}

However, a completely different situation appears in a nonmagnetizable
plasma. At high enough frequencies, or if the magnetic susceptibility
$\chi$ is small and the plasma has a long relaxation time, the magnetic
moments of particles in the plasma cannot respond to the time variation
of an electromagnetic wave in time. In this case, as was argued by
Landau and Lifshits, the magnetization loses its physical meaning,
and we need to treat this problem in an alternative approach by taking
the magnetic permeability to the vacuum value, $\mu=1$. Then the
effective permeability comes from the spatial part of the generalized
permittivity $\tilde{\epsilon}$. By expanding $\Pi_{T}(\omega,k)$
in powers of $k$ at $k=0$ \begin{equation}
\Pi_{T}(\omega,k)=m_{D}^{2}\left(\frac{1}{3}+\frac{1}{15}\frac{k^{2}}{\omega^{2}}\right)+\mathcal{O}(k^{4}),\end{equation}
we then obtain from Eq. (\ref{eq:unmagnetizable_epsilon_mu}), \begin{eqnarray}
\epsilon(\omega) & = & 1-\frac{m_{D}^{2}}{3\omega^{2}},\;\mu(\omega)=\frac{1}{1+\frac{m_{D}^{2}}{15\omega^{2}}}.\end{eqnarray}
It is obvious that $\epsilon$ is the same as in the magnetizable
case. But the effective magnetic permeability $\mu$ is always positive,
so the NRI is absent. A complete screening gap up to $\omega_{p}$
is present, only the mode with frequencies higher than $\omega_{p}$
can propagate. The dispersion relation is the same as the Fig. \ref{fig:disp-m},
but $n$ below the plasma frequency is purely imaginary ($n^{2}<0$)
and hence all modes are damped, see Fig. \ref{fig:ri-unm}. 

\subsection{A criterion for magnetizable and nonmagnetizable plasma}

As we can see from the above discussions, the magnetic response is
essential in justification of the QGP as a magnetizable or nonmagnetizable
medium. The core question is then: whether the magnetization density
has physical meaning for the plasma. To this end we look at the induced
macroscopic current density $\mathbf{J}$ from not only the magnetization
$\mathbf{M}=(\mathbf{B}-\mathbf{H})/4\pi$ but also the dielectric
polarization $\mathbf{P}=(\mathbf{D}-\mathbf{E})/4\pi$,

\begin{equation}
\mathbf{J}=\nabla\times\mathbf{M}+\frac{\partial\mathbf{P}}{\partial t},\end{equation}
which can be derived directly from the Maxwell equations

\begin{eqnarray}
\nabla\times\mathbf{B} & = & \mathbf{J}+\frac{\partial\mathbf{E}}{\partial t},\nonumber \\
\nabla\times\mathbf{H} & = & \frac{\partial\mathbf{D}}{\partial t}.\end{eqnarray}
We can compare the contribution from the electric and magnetic sector
to the induced current and see if the concept of the magnetization
still works. When the current is dominated by the magnetization, the
following condition must be satisfied\begin{equation}
\left|\nabla\times\mathbf{M}\right|\gg\left|\frac{\partial\mathbf{P}}{\partial t}\right|,\end{equation}
which is equivalent to\begin{equation}
R(\omega)=\left|\frac{\epsilon(\omega)\left(\mu(\omega)-1\right)}{\epsilon(\omega)-1}\right|\gg1.\label{eq:condition}\end{equation}
We can use Eqs. (\ref{eq:whole_dispersion_mag},\ref{eq:epsilon_mu_HTL_mag})
to test if $R(\omega)\gg1$ holds or not. If it does, the magnetization
density makes a dominant contribution to the induced current density,
so the plasma is magnetizable and Eqs. (\ref{eq:whole_dispersion_mag},\ref{eq:epsilon_mu_HTL_mag})
apply. If this condition is violated, the magnetization density is
negligible and the plasma is an nonmagnetizable medium. In this case, 
$\mu(\omega)$ in Eqs. (\ref{eq:whole_dispersion_mag},\ref{eq:epsilon_mu_HTL_mag})
loses its normal meaning and is not applicable. Thus one has to implement
the Landau-Lifshits scenario. Therefore, Eq. (\ref{eq:condition})
can be regarded as a criterion to judge if a plasma is magnetizable
or nonmagnetizable. 

\begin{figure}
\includegraphics[scale=0.5]{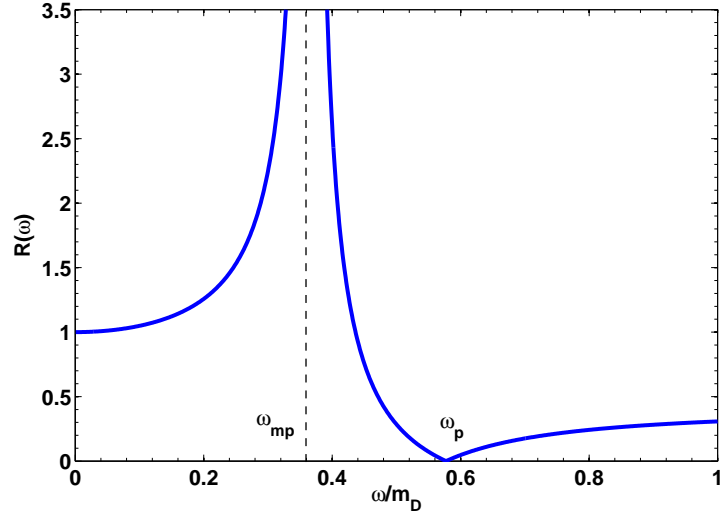}

\caption{\label{fig:R}The ratio $R(\omega)$ as a function of $\omega$ for
a fixed and small value of $k$. $R(\omega)\gg1$ is valid in the
range $\omega\lesssim\omega_{mp}$ where the refractive index becomes
negative.}

\end{figure}

In our study, as shown in Fig. \ref{fig:R}, the criterion (\ref{eq:condition})
is valid near the pole $\omega_{mp}$, below which the refractive index
becomes negative. So near the pole $\omega_{mp}$ the plasma is magnetizable,
otherwise it is nonmagnetizable. 

Although the dispersion relation shown in Fig. \ref{fig:disp-m} seems
to be the same in magnetizable and nonmagnetizable plasmas, completely
different behaviors of the RI occur below $\omega_{p}$. The main
difference is the existence of a pole in the magnetizable plasma corresponding
to a resonance. In this region, the phase velocity is slowed down
and vanishing at the pole, so the thermal velocity is much greater
than the phase velocity and the plasma becomes strongly anisotropic,
which is beyond the scope of this paper. 

Our approach is based on the HTL perturbation, which is quite different
from the holographic approach. The dispersion relation behaves differently
in two approaches. In our approach there is a frequency range with
$n^{2}<0$ corresponding to a forbidden band for electromagnetic waves
in between the negative and positive refraction region, while in the
holographic theory the dispersion relation is continuous. This distinction
is closely related to the effectiveness of the notion of quasiparticles.
In addition, there is a pole at $\omega_{mp}$ in the HTL approach
which makes $n^{2}$ and $n_{DL}$ singular. Such a property is possibly
a feature for the magnetizable plasma in a weakly coupled system.
In contrast, in the holographic theory for the nonmagnetizable and
strongly coupled plasma, there is no such a pole.

\section{Anisotropic quark matter}

\label{sec:anisotropy}In Sec. \ref{sec:htl} we have considered an
isotropic quark matter. The quark momentum distribution is anisotropic
at early time in noncentral heavy ion collisions. A proper way of
generalizing to an anisotropic case is to derive the self-energy using
the kinetic theory. Here we consider the quark distribution function
$f_{a}(p)$ for $a=u,d,s,\bar{u},\bar{d},\bar{s}$, which obey the
Boltzmann equation; the self-energy can be expressed by $f_{a}(p)$
\cite{Romatschke:2003ms,Romatschke:2004jh} \begin{equation}
\Pi_{\mu\nu}(K)=-2e^{2}N_{c}\sum_{a}Q_{a}^{2}\int\frac{d^{3}p}{(2\pi)^{3}}V_{\mu}\frac{\partial f_{a}(p)}{\partial P_{i}}\left(g_{\nu i}-\frac{V_{\nu}K_{i}}{K\cdot V+i\eta}\right)\label{eq:pi-munu1}\end{equation}
 where the four-velocity is defined by $V^{\mu}=P^{\mu}/p=(1,\mathbf{v})=(1,\mathbf{p}/p)$
and $f_{a}(p)=1/(e^{\beta p-\beta\mu_{a}}+1)$. Here the summation
for $i$ is over spatial components. Note that $\Pi_{\mu\nu}(K)$
has a different sign in our convention from Refs. \cite{Romatschke:2003ms,Romatschke:2004jh}.
To generalize Eq. (\ref{eq:pi-munu1}) to anisotropic case, we can
make replacement in distribution function $f(p)$ \begin{equation}
p\rightarrow\tilde{p}=\left[p^{2}+\xi(\mathbf{p}\cdot\mathbf{r})^{2}\right]^{1/2},\end{equation}
where $\xi$ parametrizes the strength of the anisotropy: a positive/negative
value of $\xi$ corresponds to a contraction/stretching of the isotropic
distribution function along $\mathbf{r}$. Then the self-energy becomes
\begin{eqnarray}
\Pi_{\mu\nu}(K) & = & -2e^{2}N_{c}\sum_{a}Q_{a}^{2}\int\frac{d^{3}p}{(2\pi)^{3}}V_{\mu}\frac{\partial f_{a}(\tilde{p})}{\partial P_{i}}\left(g_{\nu i}-\frac{V_{\nu}K_{i}}{K\cdot V+i\eta}\right)\nonumber \\
 & = & -m_{D}^{2}\int\frac{d\Omega}{4\pi}\frac{\mathbf{v}_{i}+\xi(\mathbf{v}\cdot\mathbf{r})\mathbf{r}_{i}}{[1+\xi(\mathbf{v}\cdot\mathbf{r})^{2}]^{2}}V_{\mu}\left(g_{\nu i}-\frac{V_{\nu}K_{i}}{K\cdot V+i\eta}\right).\end{eqnarray}
 The spatial component $\Pi_{ij}$ is then \begin{eqnarray}
\Pi_{jl} & = & -m_{D}^{2}\int\frac{d\Omega}{4\pi}\frac{\mathbf{v}_{i}+\xi(\mathbf{v}\cdot\mathbf{r})\mathbf{r}_{i}}{[1+\xi(\mathbf{v}\cdot\mathbf{r})^{2}]^{2}}\mathbf{v}_{j}\left(\delta_{li}+\frac{\mathbf{v}_{l}\mathbf{k}_{i}}{K\cdot V+i\eta}\right).\end{eqnarray}
 The four structure functions in Eq. (\ref{eq:PQCD}) can be extracted
by the following contractions:

\begin{eqnarray}
\alpha & = & (P_{\mu\nu}-C_{\mu\nu})\Pi^{\mu\nu},\nonumber \\
\beta & = & Q_{\mu\nu}\Pi^{\mu\nu},\nonumber \\
\gamma & = & (2C_{\mu\nu}-P_{\mu\nu})\Pi^{\mu\nu},\nonumber \\
\delta & = & -\frac{1}{2}\frac{\omega^{2}}{K^{2}k^{2}\tilde{r}^{2}}G_{\mu\nu}\Pi^{\mu\nu}.\end{eqnarray}
We assume that the magnitude of the anisotropy is small so that we
can perform an expansion in powers of $\xi$ for $\Pi_{jl}$, \begin{eqnarray}
\Pi_{jl}(K) & = & \Pi_{jl}^{(HTL)}(K)+\xi m_{D}^{2}\int\frac{d\Omega}{4\pi}[(\mathbf{v}\cdot\mathbf{r})\mathbf{r}_{i}-2(\mathbf{v}\cdot\mathbf{r})^{2}\mathbf{v}_{i}]\mathbf{v}_{j}\nonumber \\
 &  & \times\left(\delta_{li}+\frac{\mathbf{v}_{l}\mathbf{k}_{i}}{K\cdot V+i\eta}\right)+O(\xi^{2}),\end{eqnarray}
 where the HTL self-energy is given by, \begin{equation}
\Pi_{jl}^{(HTL)}(K)=-m_{D}^{2}k_{0}\int\frac{d\Omega}{4\pi}\frac{\mathbf{v}_{j}\mathbf{v}_{l}}{k_{0}-\mathbf{v}\cdot\mathbf{k}+i\eta}.\end{equation}

Up to $O(\xi)$ the structure functions relevant to the refractive
index of transverse modes can be obtained, \begin{eqnarray}
\alpha & = & \Pi_{T}(z)+\xi\left\{ \frac{z^{2}}{12}\left(3+5\cos2\theta\right)m_{D}^{2}-\frac{1}{6}\left(1+\cos2\theta\right)m_{D}^{2}\right.\nonumber \\
 &  & \left.+\frac{1}{4}\Pi_{T}(z)\left[(1+3\cos2\theta)-z^{2}(3+5\cos2\theta)\right]\right\} ,\\
\beta & = & \Pi_{L}(z)+\xi\biggl\{\frac{1}{6}(z^{2}-1)(1+3\cos2\theta)m_{D}^{2}+\Pi_{L}(z)\nonumber \\
 &  & \times\left(\cos2\theta-\frac{z^{2}}{2}(1+3\cos2\theta)\right)\biggr\},\\
\gamma & = & \frac{\xi}{3}[3\Pi_{T}(z)-m_{D}^{2}](z^{2}-1)\sin^{2}\theta,\end{eqnarray}
where $\Pi_{T}(z)$ and $\Pi_{L}(z)$ is the standard HTL result in
Eq. (\ref{eq:HTL}), $z=\omega/k$, and $\theta$ is the angle between
the direction of the wave vector of the light and the anisotropy,
$\cos\theta=\mathbf{\mathbf{k}}\cdot\mathbf{r}$. In the limit $\xi\rightarrow0$,
the above structure functions $\alpha$ and $\beta$ reduce to the
isotropic HTL case, while $\gamma$ and $\delta$ vanish. 

First, we focus on the magnetizable case. We diagonalize $\epsilon_{ij}$
in Eq.(\ref{eq:epsilon_mag}) in the plane perpendicular to $\mathbf{k}$,
in small $k$ approximation, we get \begin{eqnarray}
\epsilon_{L}^{(mag)}(\omega) & = & 1-\frac{m_{D}^{2}}{3\omega^{2}}+\xi\frac{m_{D}^{2}}{10\omega^{2}}\left(1-\frac{1}{3}\cos2\theta\right)+\mathcal{O}(k^{2},\xi),\\
\epsilon_{R}^{(mag)}(\omega) & = & 1-\frac{m_{D}^{2}}{3\omega^{2}}+\xi\frac{m_{D}^{2}}{15\omega^{2}}+\mathcal{O}(k^{2},\xi).\end{eqnarray}
However, in an anisotropic medium, the spatial dispersion becomes
important, as one can see in the following that the term with $k$
is always accompanied by $\xi$, so small $k$ expansion has to be
treated with more care. We have assumed $\xi\ll1$ and $k\ll1$, if
we further assume $\xi/k^{2}$ is small compared with the leading
term $1-\frac{2m_{D}^{2}}{15\omega^{2}}$, the expansion can lead
to \begin{equation}
\mu^{(mag)}(\omega,k)=\frac{1}{1-\frac{2m_{D}^{2}}{15\omega^{2}}-\xi\frac{m_{D}^{2}}{15k^{2}}\cos^{2}\theta+\mathcal{O}(k^{2},\xi)}.\end{equation}
If the ratio $\xi/k^{2}$ is not small, it becomes a singular term
of $\mathcal{O}(1/k^{2})$, the expansion in small $k$ fails, and
a more rigorous consideration is required by solving the full version
of the dispersion relation. In the nonmagnetizable case, we have\begin{align}
\epsilon_{L}^{(non)}(\omega) & =1-\frac{m_{D}^{2}}{3\omega^{2}}+\xi\frac{2m_{D}^{2}}{15\omega^{2}},\\
\epsilon_{R}^{(non)}(\omega) & =1-\frac{m_{D}^{2}}{3\omega^{2}}+\xi\frac{m_{D}^{2}}{10\omega^{2}}\left(1+\frac{1}{3}\cos2\theta\right),\end{align} 
and\begin{align}
\mu_{L}^{(non)}(\omega) & =\frac{1}{1+\frac{m_{D}^{2}}{15\omega^{2}}-\xi\frac{m_{D}^{2}}{70\omega^{2}}\left(\frac{1}{3}-\cos2\theta\right)},\\
\mu_{R}^{(non)}(\omega) & =\frac{1}{1+\frac{m_{D}^{2}}{15\omega^{2}}-\xi\frac{m_{D}^{2}}{6\omega^{2}}\left(\frac{1}{7}-\frac{1}{5}\cos2\theta\right)}.\end{align}
In this section, the anisotropy is just a small correction to the
isotropic result, the plasma frequency $\omega_{p}$ and pole $\omega_{mp}$
change a little by extra parameters $\xi$ and $\theta$ introduced
from anisotropy.

\section{Summary and conclusion}

\label{sec:conclusion}We study the electromagnetic wave properties
of the QGP within the HTL perturbation theory. The electric permittivity,
the magnetic permeability and then the optical refractive index are
calculated in magnetizable and nonmagnetizable plasmas. The optical
properties of these two types of plasmas behave differently at low
frequencies due to different definitions of the magnetic permeability
$\mu$. The plasma is magnetizable if $R(\omega)=\left|\frac{\epsilon(\omega)\left(\mu(\omega)-1\right)}{\epsilon(\omega)-1}\right|\gg1$,
while it is nonmagnetizable if this condition does not hold. 

In the magnetizable plasma, $\mu$ has the normal physical meaning,
but it fails to make physical sense in the nonmagnetizable plasma.
In the magnetizable plasma, $\mu$ has a pole at $\omega_{mp}$ below
the plasma frequency $\omega_{p}$. When $k<\omega<\omega_{mp}$,
both $n$ and $n_{DL}$ become negative but no propagating modes exist.
A frequency forbidden band or a gap with imaginary $n$ or $n^{2}<0$
is in the range $\omega_{mp}<\omega<\omega_{p}$, where electromagnetic
waves are damped. For $\omega<k$, $n$ becomes complex and the medium
is dissipative. In the nonmagnetizable plasma, both $\mu$ and $n_{DL}$
are non-negative. The damped region or the gap is in the range $\omega<\omega_{p}$. 

In contrast, the negative refraction is present for the nonmagnetizable
plasma in the strongly coupled plasma in the holographic description,
where the gap is not significant and the refractive index smoothly
connects the negative to positive refraction region. This implies
that the medium is almost transparent to the light at all frequencies.
However, the negative refraction is absent in the nonmagnetizable
plasma in our approach. For the magnetizable plasma, the NR occurs
in the region $k<\omega<\omega_{mp}$ but does not support any propagating
modes. This marks the main difference of our results from the strongly
coupled plasma in the holographic approach. 

Acknowledgment: QW thanks A. Amariti, X.-H. Ge and S.-J. Sin for helpful
discussions. QW is supported in part by the National Natural Science
Foundation of China under grant 10735040.


\begin{thebibliography}{31}
\expandafter\ifx\csname natexlab\endcsname\relax\def\natexlab#1{#1}\fi
\expandafter\ifx\csname bibnamefont\endcsname\relax
  \def\bibnamefont#1{#1}\fi
\expandafter\ifx\csname bibfnamefont\endcsname\relax
  \def\bibfnamefont#1{#1}\fi
\expandafter\ifx\csname citenamefont\endcsname\relax
  \def\citenamefont#1{#1}\fi
\expandafter\ifx\csname url\endcsname\relax
  \def\url#1{\texttt{#1}}\fi
\expandafter\ifx\csname urlprefix\endcsname\relax\def\urlprefix{URL }\fi
\providecommand{\bibinfo}[2]{#2}
\providecommand{\eprint}[2][]{\url{#2}}

\bibitem[{\citenamefont{Rischke}(2004)}]{Rischke:2003mt}
\bibinfo{author}{\bibfnamefont{D.~H.} \bibnamefont{Rischke}},
  \bibinfo{journal}{Prog. Part. Nucl. Phys.} \textbf{\bibinfo{volume}{52}},
  \bibinfo{pages}{197} (\bibinfo{year}{2004}), \eprint{nucl-th/0305030}.

\bibitem[{\citenamefont{Gyulassy and McLerran}(2005)}]{Gyulassy:2004zy}
\bibinfo{author}{\bibfnamefont{M.}~\bibnamefont{Gyulassy}} \bibnamefont{and}
  \bibinfo{author}{\bibfnamefont{L.}~\bibnamefont{McLerran}},
  \bibinfo{journal}{Nucl. Phys.} \textbf{\bibinfo{volume}{A750}},
  \bibinfo{pages}{30} (\bibinfo{year}{2005}), \eprint{nucl-th/0405013}.

\bibitem[{\citenamefont{Jacobs and Wang}(2005)}]{Jacobs:2004qv}
\bibinfo{author}{\bibfnamefont{P.}~\bibnamefont{Jacobs}} \bibnamefont{and}
  \bibinfo{author}{\bibfnamefont{X.-N.} \bibnamefont{Wang}},
  \bibinfo{journal}{Prog. Part. Nucl. Phys.} \textbf{\bibinfo{volume}{54}},
  \bibinfo{pages}{443} (\bibinfo{year}{2005}), \eprint{hep-ph/0405125}.

\bibitem[{\citenamefont{Adams et~al.}(2005)}]{Adams:2005dq}
\bibinfo{author}{\bibfnamefont{J.}~\bibnamefont{Adams}} \bibnamefont{et~al.}
  (\bibinfo{collaboration}{STAR}), \bibinfo{journal}{Nucl. Phys.}
  \textbf{\bibinfo{volume}{A757}}, \bibinfo{pages}{102} (\bibinfo{year}{2005}),
  \eprint{nucl-ex/0501009}.

\bibitem[{\citenamefont{McLerran and Toimela}(1985)}]{McLerran:1984ay}
\bibinfo{author}{\bibfnamefont{L.~D.} \bibnamefont{McLerran}} \bibnamefont{and}
  \bibinfo{author}{\bibfnamefont{T.}~\bibnamefont{Toimela}},
  \bibinfo{journal}{Phys. Rev.} \textbf{\bibinfo{volume}{D31}},
  \bibinfo{pages}{545} (\bibinfo{year}{1985}).

\bibitem[{\citenamefont{Kapusta et~al.}(1991)\citenamefont{Kapusta, Lichard,
  and Seibert}}]{Kapusta:1991qp}
\bibinfo{author}{\bibfnamefont{J.~I.} \bibnamefont{Kapusta}},
  \bibinfo{author}{\bibfnamefont{P.}~\bibnamefont{Lichard}}, \bibnamefont{and}
  \bibinfo{author}{\bibfnamefont{D.}~\bibnamefont{Seibert}},
  \bibinfo{journal}{Phys. Rev.} \textbf{\bibinfo{volume}{D44}},
  \bibinfo{pages}{2774} (\bibinfo{year}{1991}).

\bibitem[{\citenamefont{Alam et~al.}(2001)\citenamefont{Alam, Sarkar, Hatsuda,
  Nayak, and Sinha}}]{Alam:2000bu}
\bibinfo{author}{\bibfnamefont{J.-e.} \bibnamefont{Alam}},
  \bibinfo{author}{\bibfnamefont{S.}~\bibnamefont{Sarkar}},
  \bibinfo{author}{\bibfnamefont{T.}~\bibnamefont{Hatsuda}},
  \bibinfo{author}{\bibfnamefont{T.~K.} \bibnamefont{Nayak}}, \bibnamefont{and}
  \bibinfo{author}{\bibfnamefont{B.}~\bibnamefont{Sinha}},
  \bibinfo{journal}{Phys. Rev.} \textbf{\bibinfo{volume}{C63}},
  \bibinfo{pages}{021901} (\bibinfo{year}{2001}), \eprint{hep-ph/0008074}.

\bibitem[{\citenamefont{Arnold et~al.}(2001{\natexlab{a}})\citenamefont{Arnold,
  Moore, and Yaffe}}]{Arnold:2001ba}
\bibinfo{author}{\bibfnamefont{P.~B.} \bibnamefont{Arnold}},
  \bibinfo{author}{\bibfnamefont{G.~D.} \bibnamefont{Moore}}, \bibnamefont{and}
  \bibinfo{author}{\bibfnamefont{L.~G.} \bibnamefont{Yaffe}},
  \bibinfo{journal}{JHEP} \textbf{\bibinfo{volume}{11}}, \bibinfo{pages}{057}
  (\bibinfo{year}{2001}{\natexlab{a}}), \eprint{hep-ph/0109064}.

\bibitem[{\citenamefont{Arnold et~al.}(2001{\natexlab{b}})\citenamefont{Arnold,
  Moore, and Yaffe}}]{Arnold:2001ms}
\bibinfo{author}{\bibfnamefont{P.~B.} \bibnamefont{Arnold}},
  \bibinfo{author}{\bibfnamefont{G.~D.} \bibnamefont{Moore}}, \bibnamefont{and}
  \bibinfo{author}{\bibfnamefont{L.~G.} \bibnamefont{Yaffe}},
  \bibinfo{journal}{JHEP} \textbf{\bibinfo{volume}{12}}, \bibinfo{pages}{009}
  (\bibinfo{year}{2001}{\natexlab{b}}), \eprint{hep-ph/0111107}.

\bibitem[{\citenamefont{Adler et~al.}(2005)}]{Adler:2005ig}
\bibinfo{author}{\bibfnamefont{S.~S.} \bibnamefont{Adler}} \bibnamefont{et~al.}
  (\bibinfo{collaboration}{PHENIX}), \bibinfo{journal}{Phys. Rev. Lett.}
  \textbf{\bibinfo{volume}{94}}, \bibinfo{pages}{232301}
  (\bibinfo{year}{2005}), \eprint{nucl-ex/0503003}.

\bibitem[{\citenamefont{Abelev et~al.}(2010)}]{Abelev:2009hx}
\bibinfo{author}{\bibfnamefont{B.~I.} \bibnamefont{Abelev}}
  \bibnamefont{et~al.} (\bibinfo{collaboration}{STAR}), \bibinfo{journal}{Phys.
  Rev.} \textbf{\bibinfo{volume}{C81}}, \bibinfo{pages}{064904}
  (\bibinfo{year}{2010}), \eprint{0912.3838}.

\bibitem[{\citenamefont{Prasad}(2011)}]{Prasad:2011mx}
\bibinfo{author}{\bibfnamefont{S.~K.} \bibnamefont{Prasad}}
  (\bibinfo{collaboration}{ALICE}), \bibinfo{journal}{Nucl. Phys. A}
  \textbf{\bibinfo{volume}{862-863}}, \bibinfo{pages}{279}
  (\bibinfo{year}{2011}), \eprint{1103.1668}.

\bibitem[{\citenamefont{V.G.Veselago}(1968)}]{Veselago:1968}
\bibinfo{author}{\bibnamefont{V.G.Veselago}}, \bibinfo{journal}{Sov.Phys.Usp.}
  \textbf{\bibinfo{volume}{10}}, \bibinfo{pages}{509} (\bibinfo{year}{1968}).

\bibitem[{\citenamefont{Shelby et~al.}(2001)\citenamefont{Shelby, Smith, and
  Schultz}}]{Shelby:2001}
\bibinfo{author}{\bibfnamefont{R.}~\bibnamefont{Shelby}},
  \bibinfo{author}{\bibfnamefont{D.}~\bibnamefont{Smith}}, \bibnamefont{and}
  \bibinfo{author}{\bibfnamefont{S.}~\bibnamefont{Schultz}},
  \bibinfo{journal}{Science} \textbf{\bibinfo{volume}{292}},
  \bibinfo{pages}{77} (\bibinfo{year}{2001}).

\bibitem[{\citenamefont{Smith et~al.}(2004)\citenamefont{Smith, Pendry, and
  Wiltshire}}]{Smith:2004}
\bibinfo{author}{\bibfnamefont{D.}~\bibnamefont{Smith}},
  \bibinfo{author}{\bibfnamefont{J.}~\bibnamefont{Pendry}}, \bibnamefont{and}
  \bibinfo{author}{\bibfnamefont{M.}~\bibnamefont{Wiltshire}},
  \bibinfo{journal}{Science} \textbf{\bibinfo{volume}{305}},
  \bibinfo{pages}{788} (\bibinfo{year}{2004}).

\bibitem[{\citenamefont{Reed et~al.}(2003)\citenamefont{Reed, Solja\ifmmode
  \check{c}\else \v{c}\fi{}i\ifmmode~\acute{c}\else \'{c}\fi{}, and
  Joannopoulos}}]{PhysRevLett.91.133901}
\bibinfo{author}{\bibfnamefont{E.~J.} \bibnamefont{Reed}},
  \bibinfo{author}{\bibfnamefont{M.}~\bibnamefont{Solja\ifmmode \check{c}\else
  \v{c}\fi{}i\ifmmode~\acute{c}\else \'{c}\fi{}}}, \bibnamefont{and}
  \bibinfo{author}{\bibfnamefont{J.~D.} \bibnamefont{Joannopoulos}},
  \bibinfo{journal}{Phys. Rev. Lett.} \textbf{\bibinfo{volume}{91}},
  \bibinfo{pages}{133901} (\bibinfo{year}{2003}).

\bibitem[{\citenamefont{Schurig et~al.}(2006)\citenamefont{Schurig, Mock,
  Justice, Cummer, Pendry, Starr, and Smith}}]{Schurig:2006}
\bibinfo{author}{\bibfnamefont{D.}~\bibnamefont{Schurig}},
  \bibinfo{author}{\bibfnamefont{J.}~\bibnamefont{Mock}},
  \bibinfo{author}{\bibfnamefont{B.}~\bibnamefont{Justice}},
  \bibinfo{author}{\bibfnamefont{S.}~\bibnamefont{Cummer}},
  \bibinfo{author}{\bibfnamefont{J.}~\bibnamefont{Pendry}},
  \bibinfo{author}{\bibfnamefont{A.}~\bibnamefont{Starr}}, \bibnamefont{and}
  \bibinfo{author}{\bibfnamefont{D.}~\bibnamefont{Smith}},
  \bibinfo{journal}{Science} \textbf{\bibinfo{volume}{314}},
  \bibinfo{pages}{977} (\bibinfo{year}{2006}).

\bibitem[{\citenamefont{Amariti
  et~al.}(2011{\natexlab{a}})\citenamefont{Amariti, Forcella, Mariotti, and
  Policastro}}]{Amariti:2010jw}
\bibinfo{author}{\bibfnamefont{A.}~\bibnamefont{Amariti}},
  \bibinfo{author}{\bibfnamefont{D.}~\bibnamefont{Forcella}},
  \bibinfo{author}{\bibfnamefont{A.}~\bibnamefont{Mariotti}}, \bibnamefont{and}
  \bibinfo{author}{\bibfnamefont{G.}~\bibnamefont{Policastro}},
  \bibinfo{journal}{JHEP} \textbf{\bibinfo{volume}{04}}, \bibinfo{pages}{036}
  (\bibinfo{year}{2011}{\natexlab{a}}), \eprint{1006.5714}.

\bibitem[{\citenamefont{Ge et~al.}(2011)\citenamefont{Ge, Jo, and
  Sin}}]{Ge:2010yc}
\bibinfo{author}{\bibfnamefont{X.-H.} \bibnamefont{Ge}},
  \bibinfo{author}{\bibfnamefont{K.}~\bibnamefont{Jo}}, \bibnamefont{and}
  \bibinfo{author}{\bibfnamefont{S.-J.} \bibnamefont{Sin}},
  \bibinfo{journal}{JHEP} \textbf{\bibinfo{volume}{03}}, \bibinfo{pages}{104}
  (\bibinfo{year}{2011}), \eprint{1012.2515}.

\bibitem[{\citenamefont{Gao and Zhang}(2010)}]{Gao:2010ie}
\bibinfo{author}{\bibfnamefont{X.}~\bibnamefont{Gao}} \bibnamefont{and}
  \bibinfo{author}{\bibfnamefont{H.-b.} \bibnamefont{Zhang}},
  \bibinfo{journal}{JHEP} \textbf{\bibinfo{volume}{08}}, \bibinfo{pages}{075}
  (\bibinfo{year}{2010}), \eprint{1008.0720}.

\bibitem[{\citenamefont{Bigazzi et~al.}(2011)\citenamefont{Bigazzi, Cotrone,
  Mas, Mayerson, and Tarrio}}]{Bigazzi:2011it}
\bibinfo{author}{\bibfnamefont{F.}~\bibnamefont{Bigazzi}},
  \bibinfo{author}{\bibfnamefont{A.~L.} \bibnamefont{Cotrone}},
  \bibinfo{author}{\bibfnamefont{J.}~\bibnamefont{Mas}},
  \bibinfo{author}{\bibfnamefont{D.}~\bibnamefont{Mayerson}}, \bibnamefont{and}
  \bibinfo{author}{\bibfnamefont{J.}~\bibnamefont{Tarrio}},
  \bibinfo{journal}{JHEP} \textbf{\bibinfo{volume}{04}}, \bibinfo{pages}{060}
  (\bibinfo{year}{2011}), \eprint{1101.3560}.

\bibitem[{\citenamefont{Amariti
  et~al.}(2011{\natexlab{b}})\citenamefont{Amariti, Forcella, Mariotti, and
  Siani}}]{Amariti:2011dm}
\bibinfo{author}{\bibfnamefont{A.}~\bibnamefont{Amariti}},
  \bibinfo{author}{\bibfnamefont{D.}~\bibnamefont{Forcella}},
  \bibinfo{author}{\bibfnamefont{A.}~\bibnamefont{Mariotti}}, \bibnamefont{and}
  \bibinfo{author}{\bibfnamefont{M.}~\bibnamefont{Siani}},
  \bibinfo{journal}{JHEP} \textbf{\bibinfo{volume}{10}}, \bibinfo{pages}{104}
  (\bibinfo{year}{2011}{\natexlab{b}}), \eprint{1107.1242}.

\bibitem[{\citenamefont{Amariti
  et~al.}(2011{\natexlab{c}})\citenamefont{Amariti, Forcella, and
  Mariotti}}]{Amariti:2011dj}
\bibinfo{author}{\bibfnamefont{A.}~\bibnamefont{Amariti}},
  \bibinfo{author}{\bibfnamefont{D.}~\bibnamefont{Forcella}}, \bibnamefont{and}
  \bibinfo{author}{\bibfnamefont{A.}~\bibnamefont{Mariotti}}
  (\bibinfo{year}{2011}{\natexlab{c}}), \eprint{1107.1240}.

\bibitem[{\citenamefont{L.D.Landau and E.M.Lifshitz}(1984)}]{Landau:1984}
\bibinfo{author}{\bibnamefont{L.D.Landau}} \bibnamefont{and}
  \bibinfo{author}{\bibnamefont{E.M.Lifshitz}},
  \emph{\bibinfo{title}{Electrodynamics of continuous media}}
  (\bibinfo{publisher}{Butterworth-Heinemann}, \bibinfo{year}{1984}).

\bibitem[{\citenamefont{Agranovich et~al.}(2004)\citenamefont{Agranovich, Shen,
  Baughman, and Zakhidov}}]{PhysRevB.69.165112}
\bibinfo{author}{\bibfnamefont{V.~M.} \bibnamefont{Agranovich}},
  \bibinfo{author}{\bibfnamefont{Y.~R.} \bibnamefont{Shen}},
  \bibinfo{author}{\bibfnamefont{R.~H.} \bibnamefont{Baughman}},
  \bibnamefont{and} \bibinfo{author}{\bibfnamefont{A.~A.}
  \bibnamefont{Zakhidov}}, \bibinfo{journal}{Phys. Rev. B}
  \textbf{\bibinfo{volume}{69}}, \bibinfo{pages}{165112}
  (\bibinfo{year}{2004}).

\bibitem[{\citenamefont{R.A.Depine and A.Lakhtakia}(2004)}]{DL}
\bibinfo{author}{\bibnamefont{R.A.Depine}} \bibnamefont{and}
  \bibinfo{author}{\bibnamefont{A.Lakhtakia}}, \bibinfo{journal}{Microwave and
  Optical Technology Letters} \textbf{\bibinfo{volume}{41}},
  \bibinfo{pages}{315} (\bibinfo{year}{2004}).

\bibitem[{\citenamefont{Weldon}(1982)}]{Weldon:1982aq}
\bibinfo{author}{\bibfnamefont{H.~A.} \bibnamefont{Weldon}},
  \bibinfo{journal}{Phys. Rev.} \textbf{\bibinfo{volume}{D26}},
  \bibinfo{pages}{1394} (\bibinfo{year}{1982}).

\bibitem[{\citenamefont{Pisarski}(1989)}]{Pisarski:1988vd}
\bibinfo{author}{\bibfnamefont{R.~D.} \bibnamefont{Pisarski}},
  \bibinfo{journal}{Phys. Rev. Lett.} \textbf{\bibinfo{volume}{63}},
  \bibinfo{pages}{1129} (\bibinfo{year}{1989}).

\bibitem[{\citenamefont{Bellac}(1996)}]{Bellac:1996}
\bibinfo{author}{\bibfnamefont{M.~L.} \bibnamefont{Bellac}},
  \bibinfo{journal}{Cambridge University Press}  (\bibinfo{year}{1996}).

\bibitem[{\citenamefont{Romatschke and Strickland}(2003)}]{Romatschke:2003ms}
\bibinfo{author}{\bibfnamefont{P.}~\bibnamefont{Romatschke}} \bibnamefont{and}
  \bibinfo{author}{\bibfnamefont{M.}~\bibnamefont{Strickland}},
  \bibinfo{journal}{Phys. Rev.} \textbf{\bibinfo{volume}{D68}},
  \bibinfo{pages}{036004} (\bibinfo{year}{2003}), \eprint{hep-ph/0304092}.

\bibitem[{\citenamefont{Romatschke and Strickland}(2004)}]{Romatschke:2004jh}
\bibinfo{author}{\bibfnamefont{P.}~\bibnamefont{Romatschke}} \bibnamefont{and}
  \bibinfo{author}{\bibfnamefont{M.}~\bibnamefont{Strickland}},
  \bibinfo{journal}{Phys. Rev.} \textbf{\bibinfo{volume}{D70}},
  \bibinfo{pages}{116006} (\bibinfo{year}{2004}), \eprint{hep-ph/0406188}.

\end{thebibliography}
\end{document}